\documentclass[conference]{IEEEtran}
\IEEEoverridecommandlockouts
\usepackage{cite}
\usepackage{amsmath,amssymb,amsfonts}
\usepackage{algorithmic}
\usepackage[caption=false, font=footnotesize]{subfig}
\usepackage{graphicx}
\usepackage{textcomp}
\usepackage{url}
\def\BibTeX{{\rm B\kern-.05em{\sc i\kern-.025em b}\kern-.08em
    T\kern-.1667em\lower.7ex\hbox{E}\kern-.125emX}}
\begin{document}

\title{Real-Time Testbed for Diversity in Powerline and Wireless Smart Grid Communications\\
\thanks{This work was supported by the Semiconductor Research Corporation (SRC) under SRC GRC Task ID 1836.133 with industry liaisons Texas Instruments and NXP Semiconductors through the Texas Analog Center of Excellence at The University of Texas at Dallas.}
}

\author{\IEEEauthorblockN{Junmo Sung and Brian L. Evans}
\IEEEauthorblockA{\textit{Wireless Networking and Communications Group} \\
\textit{The University of Texas at Austin}, Austin, TX USA \\
junmo.sung@utexas.edu, bevans@ece.utexas.edu}
}

\maketitle

\begin{abstract}
Two-way communication is a key feature in a smart grid. It is enabled by either powerline communication or wireless communication technologies. Utilizing both technologies can potentially enhance communication reliability, and many diversity combining schemes have been proposed. In this paper, we propose a flexible real-time testbed to evaluate diversity combining schemes over physical channels. The testbed provides essential parts of physical layers on which both powerline and wireless communications operate. 
The contributions of this paper are 
1) design and implementation of a real-time testbed for diversity of simultaneous powerline and wireless communications, 
2) release of the setup information and complete source code for the testbed, and 
3) performance evaluation of maximal ratio combining (MRC) on the testbed. 
As initial results, we show that performance of MRC from measurements obtained on the testbed over physical channels is very close to that in simulations in various test cases under a controlled laboratory environment.
\end{abstract}

\section{Introduction} 
\label{sec:introduction}
People living in modern society rely on electronic devices and appliances more than ever. Put differently, our society relies on electricity more than any other in human history. Power outages, thus, degrade the quality of life and cause damage to businesses. Power outages cost the U.S. more than \$100B per year \cite{doeSGintro}. Furthermore aging infrastructure will make it challenging to meet the growing power demand. Replacing aging coal plants, which generate about one third of the electricity in the U.S. \cite{eia2016monthly}, will require an estimated \$560B by 2030. In 2015, 94 coal plants with an average age of 54 years were retired \cite{eia2016preliminary}, which amounts to nearly 5\% of total U.S. coal capacity. In traditional power grids, the unidirectional flow of information and electricity hinders precise demand prediction and integration of renewable and other energy sources. 

Smart grids which have seen initial trials by several utilities \cite{doeSmartGridSystemReport,eiaSGCaseStudies} are anticipated to be solutions for managing the growing demand by distributing energy in an unconventional and smart manner \cite{5357331,5406152,6011696}. As an integration of communication and electricity distribution networks, information conveyed through the communication network is exploited to improve reliability and efficiency of the distribution system. Smart meters, thus, are expected to play a key role in smart grid communication \cite{6520030}. They are capable of not only measuring power consumption at users' end but also exchanging information such as measured data, meter status, and control commands between smart meters and a data concentrator. 
\begin{figure}[!t]
  \centering
  \includegraphics[width=3.35in]{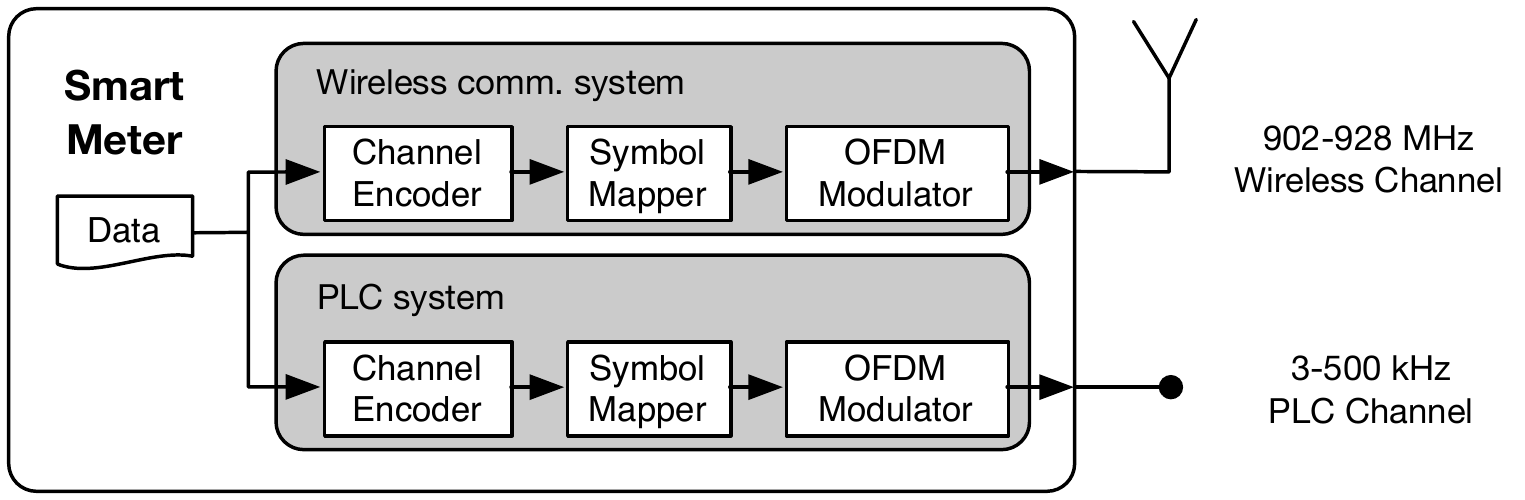}
  \includegraphics[width=3.35in]{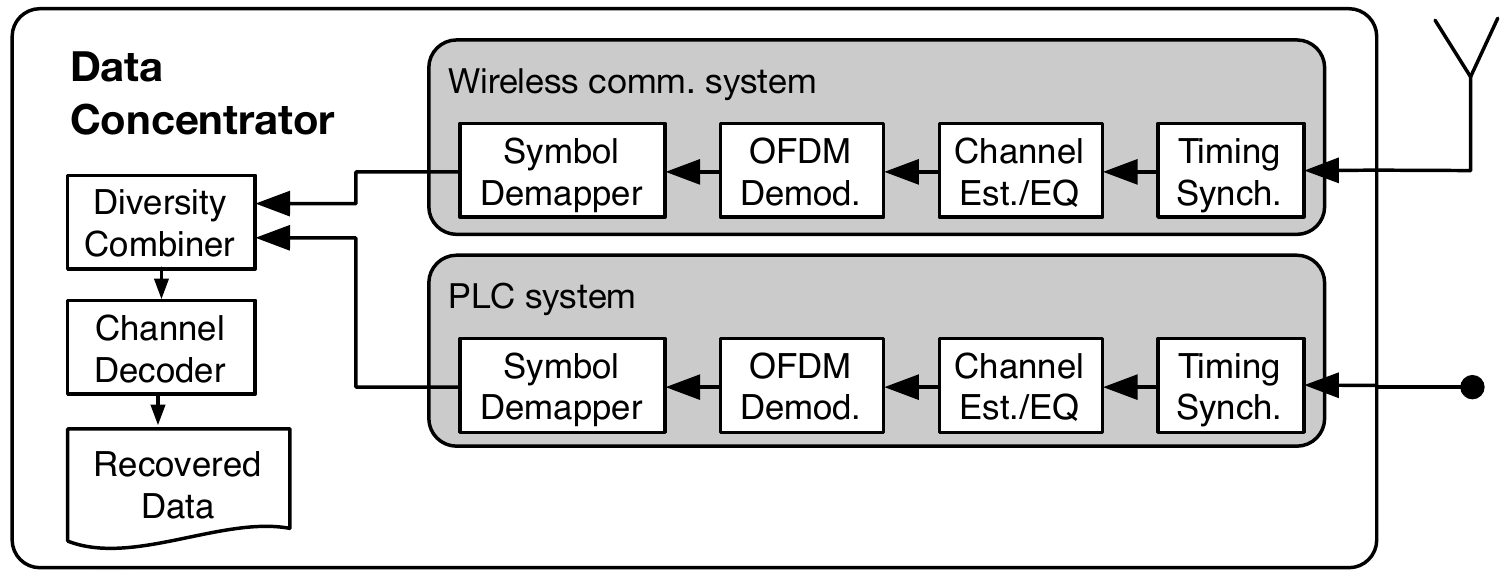}
  \caption{Diagram of PLC/wireless diversity system}
  \label{fig:system_diagram}
\end{figure}

This two-way communication is being enabled by either powerline communication (PLC) or wireless communication technologies. For powerline channels, several indoor and outdoor PLC standards such as IEEE 1901.2, ITU-T G.hn/hnem, G3, etc. were developed. There are two major contenders for the wireless communication standard in smart grid operating in the unlicensed sub-1GHz ISM band: IEEE 802.15.4g and IEEE 802.11ah. Power lines experience more complicated noise and interference, which result in modeling and mitigation techniques that are different from those for the wireless channels \cite{6279590,6288568}. Smart meters are designed to use either PLC or wireless communications or can switch between them, but do not make simultaneous use of both.   
 
One promising way to enhance communication reliability is to utilize the two independent communication technologies available for the smart grid \cite{lai2012using,Sayed:2014cq,7436327,7436267}. Fig.~\ref{fig:system_diagram} illustrates an overall diagram of a physical layer of a PLC and wireless diversity system. Theoretical studies show that diversity schemes help the network to become more reliable. 
Diversity combining schemes need to be implemented and tested in order to verify functionality and evaluate performance under realistic conditions.

Extensive surveys on smart grid testbeds were recently conducted \cite{7593427,7740849}, and a variety of targeted research areas of the testbeds include cyber security, power grid management, energy efficiency and communications. To the best of authors' knowledge, however, there has not been such testbed implementation that provides test environment for evaluation of diversity combining over the PLC and wireless communications. A flexible real-time testbed that we present in this paper is intended for such purposes. It provides essential parts of physical layers on which simultaneous over-the-wire and over-the-air smart grid communications operate. 
The contributions of this paper are 
1) we design and implement a flexible real-time testbed for evaluation of diversity combining schemes over simultaneous powerline and wireless channels,
2) we provide details of the testbed implementation including a release of the complete source code in \cite{projectcode}, and 
3) we evaluate performance of maximal ratio combining (MRC), as an example, implemented in the real-time testbed. As initial results, we show that performance of MRC from measurements obtained on the testbed is very close to that in simulations in various test cases under a controlled laboratory environment.



\begin{table}[t!]
  \renewcommand{\arraystretch}{1}
  \caption{Noise Power Estimation Approaches}
  \label{tab:noise_power_estimation}
  \centering
  \begin{tabular}{l||l||l}
  \hline
  \bfseries Average of Estimates & \bfseries Overhead & \bfseries Complexity\\
  \hline\hline
  Instantaneous & High & High\\
  Average over time \cite{7436267} & Medium & Medium \\
  Average over time and freq. & Low & Low \\
  \hline
  \end{tabular}
\end{table}

\section{Powerline and Wireless Communication Diversity} 
\label{sec:powerline_and_wireless_communication_diversity}

Assuming a subcarrier bandwidth in orthogonal frequency devision multiplexing (OFDM) is small enough compared to the coherence bandwidth of a channel, received symbols in each subcarrier, $Y_{n,k}$, can be expressed as
\begin{align}
  Y^p_{n,k} &= H^p_{n,k}X_{n,k} + N^p_{n,k} \label{eq:y_p}, \\
  Y^w_{n,k} &= H^w_{n,k}X_{n,k} + N^w_{n,k} \label{eq:y_w},
\end{align}
where $X_{n,k}$, $H_{n,k}$ and $N_{n,k}$ are the transmitted symbol, channel impulse response and additive noise in the frequency domain, respectively, at the $k^{th}$ subcarrier in the $n^{th}$ OFDM symbol, and $(\cdot)^{p}$ and $(\cdot)^{w}$ denote PLC and wireless links, respectively. For simulations and implementation, we assume narrowband channels and Gaussian noises for both communication links to obtain initial results. 

The two communication links can be exploited for higher data rate or better reliability in the smart grid. We apply diversity in the heterogeneous communications to combine two received symbols in order to improve reliability. Among combining schemes, MRC has a good trade off in communication performance versus implementation complexity \cite{lai2012using}. A log-likelihood function for MRC can be expressed as 
\begin{align*}
  LL(x) 
  &= -\frac{|Y^p_{n,k}-H^p_{n,k}X_{n,k}|^2}{\sigma_{p,k}^2} - \frac{|Y^w_{n,k}-H^w_{n,k}X_{n,k}|^2}{\sigma_{w,k}^2},
\end{align*}
where $\sigma_{p,k}^2$ and $\sigma_{w,k}^2$ are Gaussian noise variances of powerline and wireless communications, respectively, as in \cite{lai2012using, 7436267}. As discussed in \cite{7436267}, there are several approaches on averaging estimates of noise power which are summarized in Table~\ref{tab:noise_power_estimation}. We take the approach proposed in \cite{7436267} and modify it as the testbed performs channel estimation using preambles. Testbed architecture and frame structure are described in the following section.

There are many assumptions that are taken for granted in theoretical work; however, in implementation, various techniques are to be employed to meet or relax the assumptions. Some assumptions remain in the current testbed, which are described later in this paper, and will be addressed in future development. A list of assumptions and approaches to realize or tackle each is described in Section~\ref{sec:testbed_design_and_implementation_challenges}.
\begin{table}[t!]
  \renewcommand{\arraystretch}{1}
  \caption{Parameters for common baseband transceiver for simultaneous PLC and wireless communications}
  \label{tab:system_parameters}
  \centering
  \begin{tabular}{l||l}
  \hline
  \bfseries Parameter & \bfseries Value\\
  \hline\hline
  Sample rate & 400 kHz \\
  DFT/IDFT size & 256 \\
  CP length & 64 samples \\
  Frame duration & variable \\
  Frame rate & 2.5 frames/s \\
  Active subcarriers & 36 \\
  Modulation scheme & BPSK / Differential BPSK \\
  \hline
  \end{tabular}
\end{table}

\section{Testbed Overview} 
\label{sec:testbed_overview}
Smart grid wireless communication standards of IEEE 802.11ah and IEEE 802.15.4g and powerline communication standards of G3, PRIME and IEEE P1901.2 use OFDM. Even so, these standards do not share the same system parameter values. For the sake of implementation  simplicity, we use a common baseband transceiver for both wireless and powerline communications with the parameters given in Table~\ref{tab:system_parameters}. Most parameters are chosen from IEEE P1901.2 \cite{p1901.2}. However, for more realistic implementation, we consider IEEE 802.11ah for the wireless communication system in the next phase of development. A carrier frequency for wireless communications is set in the unlicensed 900 MHz band at 920 MHz.

The purpose of the real-time PLC/wireless diversity testbed is to provide an environment for testing and evaluating various combining schemes under realistic conditions in real time. The testbed, thus, is able to gather received frames or decoded bits from both links for combining. It is also able to conduct separate frame decoding over each link to show improvement achieved with a combining scheme. An overview of the testbed platform is given in this section.
\begin{figure}[!t]
  \centering
  \includegraphics[width=7.5cm]{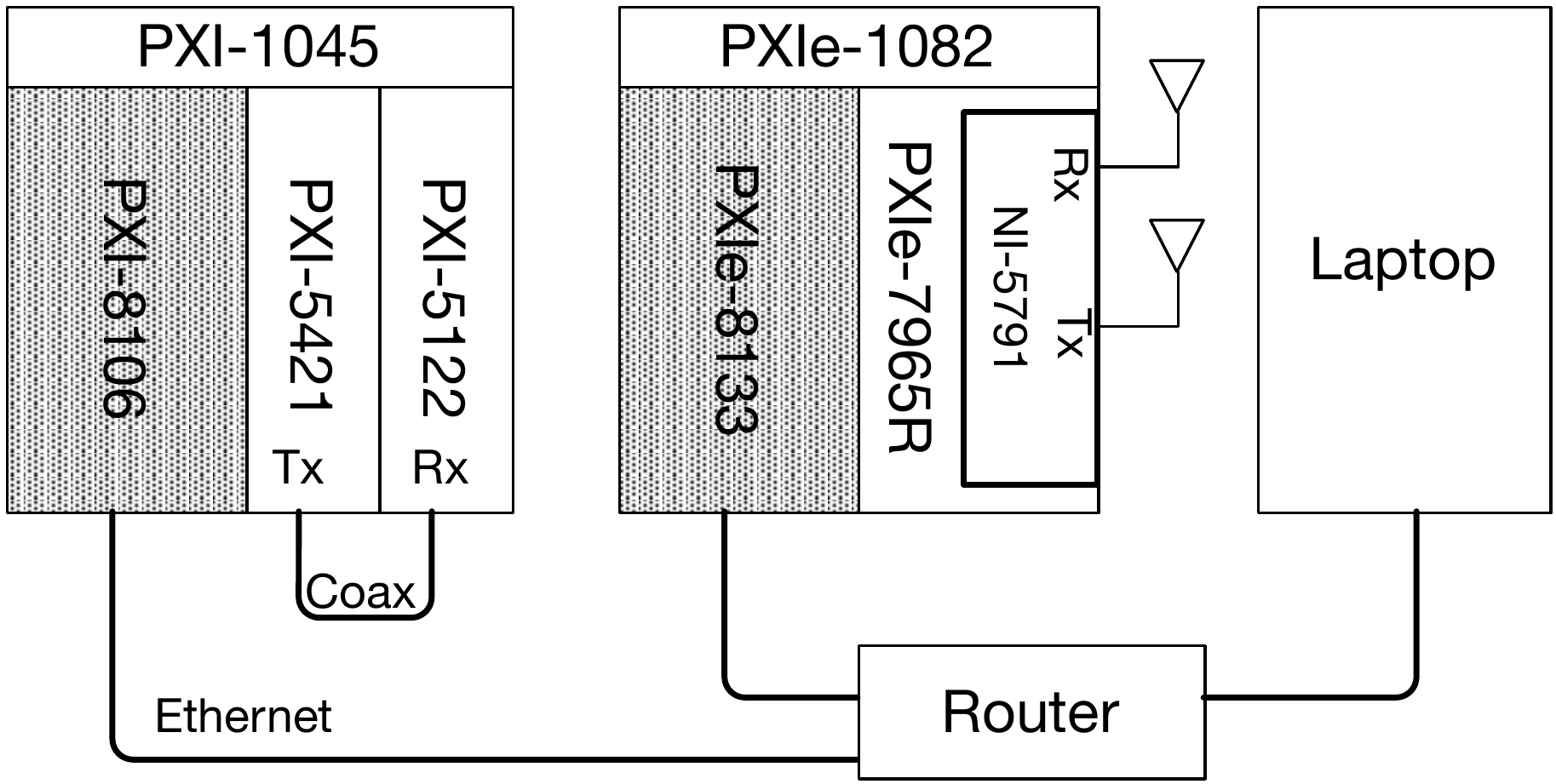}
  \caption{Hardware architecture includes two communication systems and a computer. They are interconnected via an Ethernet router.}
  \label{fig:hardware}
\end{figure}

\subsection{Hardware Architecture}
\label{subsec:hardware_archtecture}
The real-time testbed is built using products from National Instruments as shown in Fig.~\ref{fig:hardware}. Two chassis on the left and in the middle are for PLC and wireless communications, respectively. 
A PXI chassis has slots that can accommodate an x86 controller and various modules. As Fig.~\ref{fig:hardware} depicts, a PXI-1045 chassis on the left has a PXI-8106 controller, a PXI-5421 signal generator and a PXI-5122 digitizer. This chassis functions as a baseband PLC system. Similarly a PXIe-1082 chassis contains a PXIe-8133 controller, a PXIe-7965R FPGA module and an NI-5791 RF adapter module. Since the RF adapter module has both a transmit and a receive port, a unidirectional single-input single-output link can be established with a single adapter module. 

The transmit and receive antennas for the wireless system are currently spaced about one meter apart to satisfy the far-field assumption. The transmit and receive ports in the PLC system are directly connected with a short coaxial cable for testing purposes. These, however, are ready to be modified to meet research-specific demand. For example, the antennas can be placed where desired wireless interference exists, and the cable can be replaced with a powerline emulator or a real powerline. 

The two controllers and a laptop computer are interconnected via an Ethernet router so that the laptop can act as a command center and a controller. 

\begin{figure}[!t]
  \centering
  \includegraphics[width=3.4in]{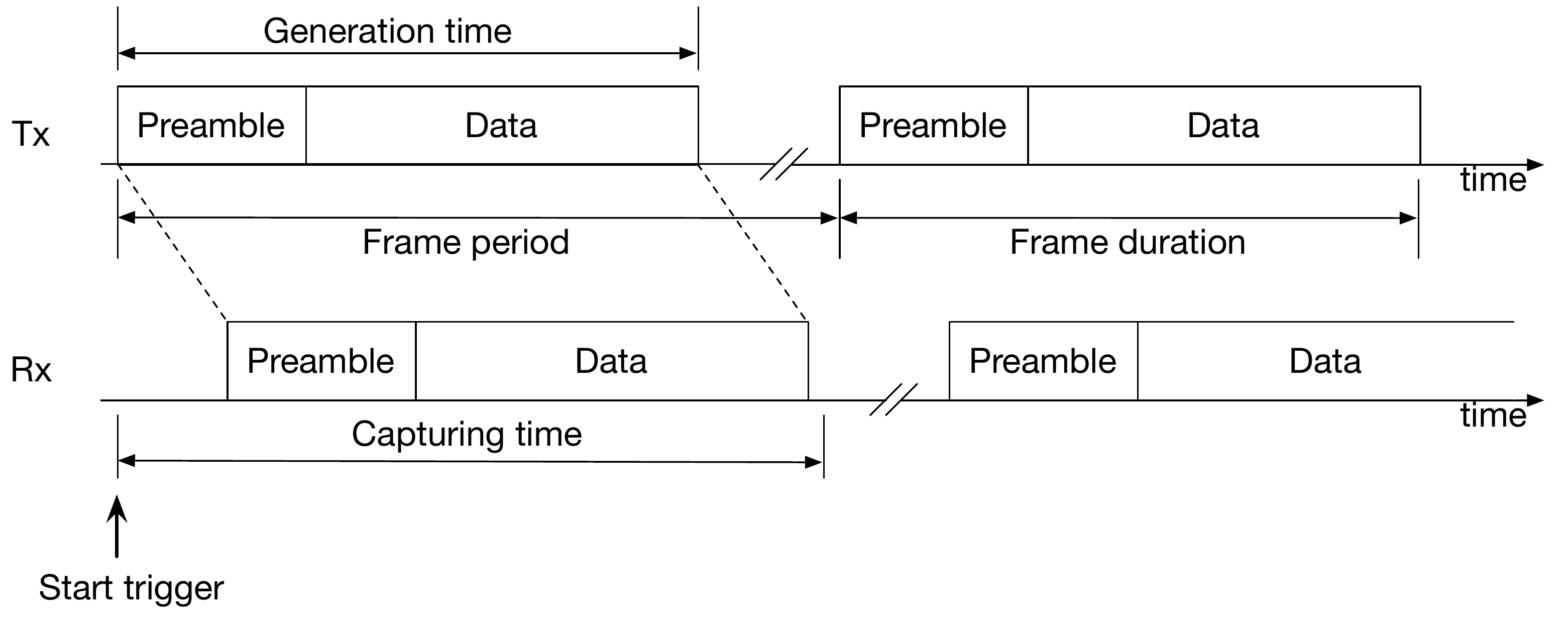}
  \caption{Frame structure and timing diagram. Start triggers are shared between a transmitter and a receiver. The receiver captures longer than an actual frame duration.}
  \label{fig:frame}
\end{figure}

\subsection{Software Architecture}
\label{subsec:software_architecture}
The two controllers in the testbed run a real-time (RT) operating system (OS), and each RTOS runs its main program. 
The controller dedicated to the PLC system runs one thread performing frame generation, bit recovery, and transmission and reception of signals with hardware. The transmitter output signal can optionally bypass the hardware and be directly sent to a receive process in case one wants to test the transceiver without channels or with simulated channels. The other controller for the wireless system also performs the same process with different hardware. These threads are configured to iterate the whole process at the same rate which determines a frame period. 

In addition, there is one more thread in the wireless system controller that carries out diversity combining. Log-likelihood ratios (LLRs) obtained from the PLC and wireless communication threads are forwarded to this thread along with a transmit bit sequence over first-in first-out (FIFO) queues and are processed in real time. 
It compares two transmit bit sequences to make sure the frames that it is trying to combine contain the same information, and then proceeds with combining. 

\subsection{Frame Structure} 
\label{sub:frame_structure}
A frame consists of a variable number of OFDM symbols. Per the IEEE P1901.2 standard \cite{p1901.2}, the first nine symbols in a frame compose a preamble which is used for channel estimation, noise estimation and timing synchronization. The number of active subcarriers is 36 in the testbed as we consider transmission in the CENELEC A frequency band (35.9375 -- 90.6250 kHz).
The following OFDM symbols -- data symbols -- contain transmitted data on the same subcarriers as the preamble.
BPSK or differential BPSK symbols made with a pseudo random bit sequence are mapped onto active subcarriers. The requested number of OFDM symbols compose the data part. Thus the frame has a variable length and is transmitted every 400 ms as shown in Fig.~\ref{fig:frame}. 

\section{Testbed Design and Implementation Challenges} 
\label{sec:testbed_design_and_implementation_challenges}
Challenges in designing and implementing the testbed are equivalent to how to deal with assumptions that are posited in theory. Assumptions we take care of in this testbed implementation are listed in Table~\ref{tab:assumptions} along with a summary of realization approaches and sections for the details.

\begin{table}[t!]
  \renewcommand{\arraystretch}{1}
  \caption{Realizing Assumptions in Communication Theory in the Testbed}
  \label{tab:assumptions}
  \centering
  \begin{tabular}{l||l||l}
  \hline
  \bfseries Assumption & \bfseries Realization & \bfseries Section\\
  \hline\hline
  A1. Channel knowledge & LS estimation & \ref{sub:channel_and_noise_estimation}\\
  A2. Noise variance knowledge & Estimation \cite{5201011} & \ref{sub:channel_and_noise_estimation}\\
  A3. Sample timing detection & Peak detection or manual & \ref{sub:sample_timing_detection}\\
  A4. Sampling frequency offset & Reference clock synch. & \ref{sub:sampling_and_carrier_frequency_offsets}\\
  A5. Carrier frequency offset & Local oscillator synch. & \ref{sub:sampling_and_carrier_frequency_offsets}\\
  A6. Frame transmission synch. & Deterministic OS & \ref{sub:frame_synchronization}\\
  \hline
  \end{tabular}
\end{table}

\subsection{Channel and Noise Estimation} 
\label{sub:channel_and_noise_estimation}
Channel and noise are intentionally generated in simulation. Hence one has direct access to parameters and instantaneous realizations of channel and noise models and takes advantage of it. The testbed, however, has to estimate parameters to exploit them in signal processing. A preamble in a frame is used for such estimation.

As with other common OFDM communication systems, the testbed employs a frequency domain least square (LS) channel estimator. There are nine OFDM symbols in the preamble, and the channel estimates obtained from the those symbols are averaged in order to reduce the effects of noise. Channel estimation takes place once at the beginning of every frame, and the estimate is used to equalize data symbols in the same frame. For further improvement in estimation, additional pilots may be embedded within data symbols or other estimation techniques may be used.

Frequency domain zero forcing (ZF) channel equalization is then performed on data symbols with the obtained channel estimate, and data bits are recovered through symbol demapping. The testbed takes advantage of channel coding and employs the convolutional encoder along with the Viterbi decoder. The symbol demapper calculates LLRs and sends them to the decoder of each system. At the same time, the LLRs are sent to the combining thread as well to perform diversity.

Noise variance and $E_b/N_0$ are estimated and averaged in real-time at the receiver with the preamble in every frame. 
The proposed noise variance estimation technique in \cite{5201011} is useful when a preamble consists of more than one identical OFDM symbol such as one used in this testbed. 

\subsection{Sample Timing Detection} 
\label{sub:sample_timing_detection}
Receivers in this testbed do not capture signals all the time, but start simultaneously with transmitters by a trigger and stop upon receiving a requested number of samples. The time of arrival of frame is unknown because of a propagation delay and a channel delay even if the transmitter and the receiver start simultaneously. For this reason, the receivers capture longer than the frame length as shown in Fig.~\ref{fig:frame}. The receivers figure out a starting point of the received frame and place the FFT window on the cyclic prefix (CP) boundary. Sample timing detection, thus, is one of the most important and earliest processes on the receiver side. To find a starting point of a frame, peaks of cross correlation between a known OFDM symbol of preamble and a received preamble in a capture are detected and used. 
To guarantee that the detected starting point falls under a CP and that the FFT window is properly located, a moderate number of sample timing advance is applied. 

Sample timing detection may not work correctly when the transmit power is too low. The testbed, therefore, provides an option to manually correct the timing. The timing offset remains unchanged as long as an antenna distance or a cable length does not change. Once a correct offset is obtained when the power is high enough, one can switch over to manual correction and use the obtained offset for low transmit power.

\subsection{Sampling and Carrier Frequency Offsets} 
\label{sub:sampling_and_carrier_frequency_offsets}
Offsets in sampling and carrier frequencies are commonly ignored in simulation as these are caused by separate oscillators in hardware. The offsets can be compensated with additional signal processing or canceled by synchronizing the oscillators. The testbed takes the synchronization approach for initial results since performance degradation due to additional signal processing can be avoided. 

A transmitter and a receiver are synchronized in two ways. The PLC system synchronizes sample clocks of the transceiver to a reference clock generated in a backplane of a PXI chassis to prevent two sample clocks from drifting away. The transmitter also sends out start triggers to the receiver to ensure they can start simultaneously. The wireless system employs the identical synchronization technique. Plus, a single local oscillator located in the adapter module is used for both up/downconversion, which results in a zero carrier frequency offset. These synchronization techniques do not need extra signal processing to compensate for impairments; however, for practicality, oscillator synchronization will be removed and compensation algorithms will be employed in the next phase of development.  

\subsection{Frame Transmission Synchronization}
\label{sub:frame_synchronization}
In order to prevent overflow of the FIFO queues in the combining thread, frame transmission of the two systems starts at the same time and transmission rates are the same. However, they operate independently without sharing any triggers or signals over a wire. The testbed, thus, takes advantage of the deterministic RTOS and uses a timed-loop in both the PLC and the wireless threads. The start, idle and stop flags are generated from the wireless system and are read by the PLC system over TCP/IP. This may introduce a delay up to several hundred milliseconds; however, once the systems start, they run at the same speed with a negligible jitter. 

\section{Initial Results} 
\label{sec:Initial_results}
A combining scheme of MRC is used in this section for testbed demonstration. As the log-likelihood function described in \cite{7436267} indicates, the LLR computed with the MRC scheme is a sum of LLRs from both communications. 

Two cases are taken into account for demonstration. In the first case, one of the links is in good shape so that all transmitted data can be recovered without an error while the other link is poor. We see that MRC can recover data even if one of the links is completely down. In the second case, $E_b/N_0$ for both systems is identical and is swept to plot BER curves. Or $E_b/N_0$ of one system, the wireless communication system in this case, is fixed and the other changes. We see that performance of MRC over physical channels is very comparable with that in simulation.

\begin{table}[t!]
  \vspace{0.1in}
  \caption{Combiner Performance}
  \label{tab:combiner_performance_one_bad_link}
  \centering
  \subfloat[Wireless link is down]{
    \renewcommand{\arraystretch}{1}
    \begin{tabular}{c||c|c|c}
    \hline
    \bfseries  & PLC & Wireless & Combiner\\
    \hline\hline
    $E_b/N_0$ & 7.30 dB & N/A & N/A\\
    BER & 0 & 5.00$\times 10^{-1}$ & 0\\
    \hline
    \end{tabular}
    }
  \hfill
  \subfloat[PLC link is down]{
    \renewcommand{\arraystretch}{1}
    \centering
    \begin{tabular}{c||c|c|c}
    \hline
    \bfseries  & PLC & Wireless & Combiner\\
    \hline\hline
    $E_b/N_0$ & N/A & 8.56 dB & N/A\\
    BER & 5.00$\times 10^{-1}$ & 0 & 0\\
    \hline
    \end{tabular}
    }
  
\end{table}

\subsection{One Link Is Down} 
\label{sub:one_link_down}
In the first test case, one link is reliable, and the other is completely down, or equivalently, a transmitter does not generate a signal at all. This scenario sounds trivial, and results look obvious. However, it is a necessary test case to examine if one communication system can function as a backup for the other. With an MRC scheme, smart grid needs not physically switch over between powerline and wireless communications.

The testbed transmits more than 200,000 frames over physical channels to measure BERs of each case. 
When one of the channels is reliable enough, we can see from Table~\ref{tab:combiner_performance_one_bad_link} that the combiner always yields zero BER proving that the MRC scheme can recover transmitted information even when the other channel is completely down. It is quite an expected result, but we stress that it is demonstrated that diversity of heterogeneous communication technologies helps to build a more reliable smart grid communication system.
\begin{figure}[!t]
  \vspace{0.1in}
  \centering
  \subfloat[]{
    \includegraphics[width=8.3cm]{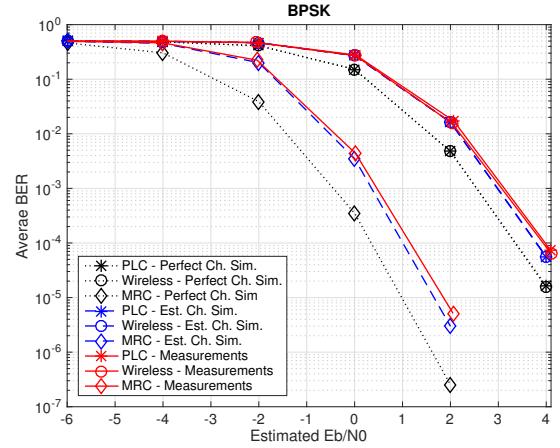}
    \label{fig:ber_curve_bpsk}
  }
  \hfill
  \subfloat[]{
    \includegraphics[width=8.3cm]{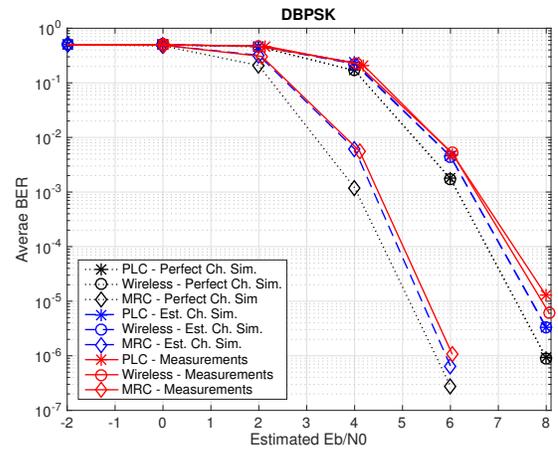}
    \label{fig:ber_curve_dbpsk}
  }
  \caption{BER vs. $E_b/N_0$ curves for (a) BPSK and (b) differential BPSK modulations}
  \label{fig:ber_curves}
\end{figure}

\begin{figure}
  \centering
  \subfloat[]{
    \includegraphics[width=8.3cm]{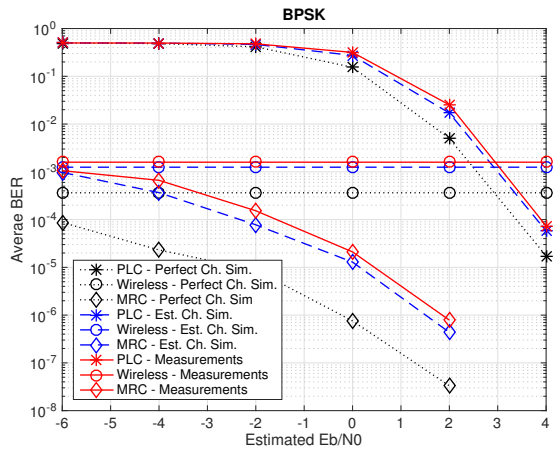}
    \label{fig:ber_curve_bpsk_fixed_wireless}
  }
  \hfill
  \subfloat[]{
      \includegraphics[width=8.3cm]{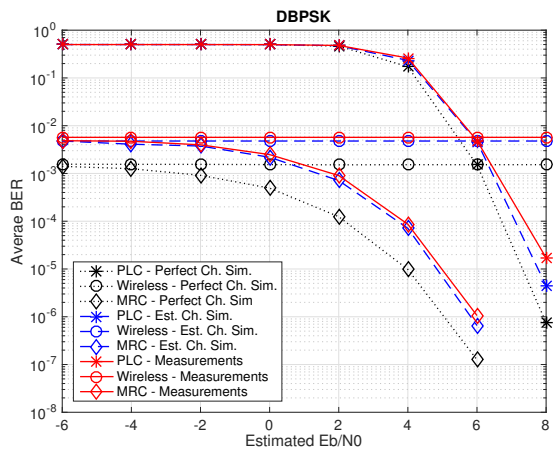}
      \label{fig:ber_curve_dbpsk_fxed_wireless}
  }
  \caption{BER vs. $E_b/N_0$ curves. $E_b/N_0$ for wireless system is fixed at (a) 3 dB for BPSK, and (b) 6 dB for differential BPSK}
  \label{fig:ber_curves_wireless_snr_fixed}  
\end{figure}

\subsection{Bit Error Rate} 
Performance of MRC measured on the testbed is compared with simulation by plotting BER curves. As shown in Figs.~\ref{fig:ber_curves} and \ref{fig:ber_curves_wireless_snr_fixed}, we compare three sets of curves : 1) perfect channel and noise variance information simulation, 2) estimated channel and noise variance simulation and 3) testbed measurements. Measurements made on the testbed are essentially based on estimation. In each set, there are three curves: PLC, wireless and combined.

Two subfigures in Fig.~\ref{fig:ber_curves} are with BPSK and differential BPSK, respectively. A set of dotted line curves is a baseline due to perfect channel and noise variance knowledge known at the receiver. A set of measurement curves cannot achieve lower BER than the other sets because of employment of hardware as well as imperfect channel and noise information. In both BPSK and differential BPSK cases, nevertheless, BER curves from measurements are very close to those in simulations using estimated channels.

Focusing on the BPSK modulation shown in Fig.~\ref{fig:ber_curve_bpsk}, an $E_b/N_0$ gain at BER of $10^{-4}$ obtained by MRC with perfect knowledge is approximately 3 dB. 
At the same BER, the measured gain on the testbed is almost 3 dB as well. In other words, it is validated with the testbed that performance of MRC in simulation can closely be achieved in practice. 

For differential BPSK shown in Fig.~\ref{fig:ber_curve_dbpsk}, an $E_b/N_0$ loss due to imperfect estimation is smaller than that in the BPSK case. That is because differential BPSK modulation does not require a channel equalizer and is not affected by channel estimation. As with BPSK, the measurements show that almost the same $E_b/N_0$ gain is achievable as simulation. Note that $E_b/N_0$ ranges are different in two subfigures in Fig.~\ref{fig:ber_curves}, and differential BPSK needs higher $E_b/N_0$ than BPSK to achieve similar performance. 

Fig.~\ref{fig:ber_curves_wireless_snr_fixed} shows BER curves as with Fig.~\ref{fig:ber_curves}, but an $E_b/N_0$ for the wireless system is fixed at 3 dB and 6 dB for BPSK and differential BPSK, respectively. With both modulations, it is seen that MRC improves BER as $E_b/N_0$ of the PLC system increases. Improvement is prominent with low $E_b/N_0$. In the $E_b/N_0$ range from -6 to -2 in Fig.~\ref{fig:ber_curve_bpsk_fixed_wireless} and -6 to 2 in Fig.~\ref{fig:ber_curve_dbpsk_fxed_wireless}, BER of the PLC stays around 0.5; however, MRC still can extract information from the poorly performing system and evidently suppresses BER. Fig.~\ref{fig:ber_curves_wireless_snr_fixed} can be alternatively interpreted that performance of the PLC can be drastically boosted with the help of the wireless system.

\section{Conclusion and Future Work}
\label{sec:conclusions}
In this paper, we proposed implementation of a flexible real-time testbed for evaluation of diversity combining schemes over powerline and wireless communications. Detailed assumptions and approaches in designing and implementing the testbed are also provided. The testbed consists of powerline and wireless OFDM communication systems using the CENELEC A band and the unlicensed sub-1GHz band, respectively, and is ready to emulate a smart grid communication network. Similar transceiver processing is employed by the two systems and is flexible enough to be modified per research demand.

We demonstrated the testbed in two different scenarios with the combining scheme of MRC. It was first shown that diversity makes smart grid communications more robust by recovering transmitted information even when one of two available links is completely down. We then plotted BER curves obtained from measurements and simulations to show that performance of MRC in practice is comparable with simulations in several test cases. The measurements also showed that performance of PLC can be drastically enhanced with the help of stable wireless communication. 

As the current testbed was designed to obtain initial results, there is plenty of room for improvement. In future work, the wireless communication system will employ one of the standards that are developed for wireless communication in smart grid such as IEEE 802.11ah. We will also add more reality to the testbed by 1) replacing a coaxial cable with a PLC channel emulator or an actual power grid, 2) generating intentional wireless interference or placing the testbed in a place with desired interference, and 3) removing clock synchronization and employing compensation algorithms. More advanced combining schemes will be tested as well.


\bibliographystyle{IEEEbib}
\bibliography{paper}

\end{document}